\documentclass[twocolumn]{aastex631}
\usepackage{natbib}
\usepackage{amsmath}
\usepackage{graphicx}
\usepackage{media9}
\usepackage{hyperref}

\usepackage{soul}

\shortauthors{Monga et al.}

\graphicspath{{./}{figures/}}

\begin{document}

\title{Driving mechanisms of solar active region geysers: The role of interacting magnetic flux}

\correspondingauthor{Aabha Monga}
\email{aabha.monga@njit.edu}

\author[0000-0002-2036-1765]{Aabha Monga}
\affiliation{Center for Solar-Terrestrial Research, New Jersey Institute of Technology, University Heights, Newark, New Jersey - 07102}
\affiliation{Institute for Space Weather Sciences, New Jersey Institute of Technology, University Heights, Newark, New Jersey - 07102}

\author[0000-0002-0197-9041]{Satoshi Inoue}
\affiliation{Center for Solar-Terrestrial Research, New Jersey Institute of Technology, University Heights, Newark, New Jersey - 07102}
\affiliation{Institute for Space Weather Sciences, New Jersey Institute of Technology, University Heights, Newark, New Jersey - 07102}

\author[0000-0002-5865-7924]{Jeongwoo Lee}
\affiliation{Center for Solar-Terrestrial Research, New Jersey Institute of Technology, University Heights, Newark, New Jersey - 07102}
\affiliation{Institute for Space Weather Sciences, New Jersey Institute of Technology, University Heights, Newark, New Jersey - 07102}

\author[0000-0002-5233-565X]{Haimin Wang}
\affiliation{Center for Solar-Terrestrial Research, New Jersey Institute of Technology, University Heights, Newark, New Jersey - 07102}
\affiliation{Institute for Space Weather Sciences, New Jersey Institute of Technology, University Heights, Newark, New Jersey - 07102}

\author[0000-0003-0975-6659]{Viggo Hansteen}
\affiliation{Lockheed Martin Solar and Astrophysics Laboratory, Palo Alto, CA 94304, USA} 
\affiliation{SETI Institute, 339 N Bernardo Ave Suite 200, Mountain View, CA 94043}
\affiliation{Rosseland Centre for Solar Physics, University of Oslo,  PO Box 1029 Blindern, 0315 Oslo, Norway}

\begin{abstract}
Active region recurrent jets are manifestations of episodic magnetic energy release processes driven by complex interactions in the lower solar atmosphere. While magnetic flux emergence and cancellation are widely recognized as key contributors to jet formation, the mechanisms behind repeated magnetic reconnection remain poorly understood. In this letter, we report a sequence of nine recurrent jets originating from active region AR 12715 during its decay phase, where the jet activity was associated with a complex distribution of fragmented magnetic flux. Non-linear force-free field (NLFFF) extrapolations reveal the presence of low-lying, current-carrying loops beneath overarching open magnetic fields near the jet footpoints. These magnetic structures were perturbed by (i) emerging flux elements and (ii) interactions between oppositely polarized moving magnetic features (MMFs). To interpret these observations, we compare them with 3D radiative MHD simulation from the Bifrost model, which reproduce jet formation driven by interacting bipolar MMFs, leading to subsequent flux cancellation in the photosphere. Our results emphasize the critical role of MMF-driven flux interactions in initiating and sustaining recurrent jet activity in active regions.

\end{abstract}

\keywords{Sun: activity,
Sun: magnetic fields,
stars: jets, 
magnetic reconnection,
magnetohydrodynamics (MHD)}

\section{Introduction}\label{Intro}
Coronal jets are collimated eruptions typically observed in extreme ultraviolet (EUV) and X-ray emissions \citep[][see references therein] {Shibata1992, Cirtain2007}, in quiet-Sun regions, coronal holes, and near active regions (ARs). It is now established that a significant fraction ($\sim$68\%) occur in the proximity of ARs and are believed to be driven by magnetic flux cancellation between opposite-polarity fields in mixed-polarity regions \citep{Shimojo1996, Shimojo2000}. These jets  make significant contributions to the heating of the solar atmosphere and transfer mass, energy, and momentum into the solar wind \citep{Paraschiv2019, Raouafi2016, Sterling2020}. A comprehensive review of the nature and implications of coronal jets is provided by \citet{Raouafi2016, Schmieder2022}.

In some cases, AR jets exhibit multiple eruptions originating from the same source region, commonly referred to as ``recurrent" or ``geyser-like" jet eruptions. \citep{Paraschiv2019, Paraschiv2020}. This behavior is generally attributed to dynamic magnetic processes, such as, flux emergence, cancellation, and/or shearing motions at the footpoint that enable repeated magnetic reconnection. Among these, magnetic flux cancellation is widely reported as the primary driver of recurrent jets \citep{Chae1999, Chen2015, Yang2023}, while other studies emphasize flux emergence as a dominant trigger at AR sites \citep{Asai2001, Guo2013, Paraschiv2020}. A combination of both emergence and cancellation processes has also been reported to generate recurrent jets at locations where opposite magnetic polarities emerge, converge, and cancel each other \citep{Jiang2007, Chifor2008, Miao2019}. Recently, \cite{Li2022} suggested a possible relationship between quasi-periodic pulsations and recurrent jet activity mediated by repetitive reconnection processes at the host AR.

The origins of reconnection-driven recurrent jets have also been explored through numerical simulations guided by observational data. \cite{Shibata1992} proposed a two-dimensional standard jet model driven by magnetic reconnection between open and closed magnetic field topologies. The closed magnetic fields were formed by bipolar magnetic structures surrounded by unipolar open magnetic fields, mimicking x-ray jets in coronal holes. This model for jet formation was further investigated by \cite{Yokoyama1995, Yokoyama1996, Miyagoshi2004}, in both 2D and 3D simulation setups. \cite{Moreno-Insertis2008} modeled interactions between tilted magnetic fields and sheared flux ropes as potential sources that trigger jet structures. \cite{Pariat2009, Pariat2010} emphasized the role of continuous twisting motions in the photosphere in generating recurrent quasi-homologous jets within a three-dimensional null-point and fan-separatrix topology. \cite{Pontin2013} also demonstrated jet formation in a region with three-dimensional spine-fan reconnection around a null point. In contrast, \cite{Moreno-Insertis2013} proposed that recurring jets resemble miniature coronal mass ejections, or blowout jets, as described by \cite{Moore2010}. Moreover, \cite{Archontis2010} simulated recurrent jet eruptions in flux emergence scenarios, where newly emerging magnetic fields within an active region perturb the pre-existing field, triggering successive magnetic reconnections and multiple jets from a common site. Furthermore, \cite{Yang2013} investigated how horizontal motions of magnetic structures in the photosphere drive jets in a coronal hole environment. \cite{Pariat2015, Pariat2016, Wyper2018} introduced the `breakout' model, where microfilament eruptions confined within open magnetic fields drive blowout jets, a mechanism also suggested by \cite{Sterling2015}. Recently, \cite{Nobrega-Siverio2023} proposed that photospheric convective motions can stress the magnetic topology of coronal bright points, leading to jet-like flows.

It is important to note that most observational studies of AR recurrent jets report the presence of small-scale, fragmented magnetic flux structures near the jet bases. These rapidly moving fragmented features, also known as moving magnetic features \cite[MMFs:][]{Harvey1973}, usually observed in the AR penumbra, propagate outward and interact with pre-existing opposite polarity magnetic fields near the host AR to produce jets \citep{Canfield1996}. MMFs move across the photosphere near ARs at higher velocities ($\sim$ 0.5-3 kms$^{-1}$) compared to background flows ($\leq$ 0.5 kms$^{-1}$), with shorter-lived MMFs typically exhibiting faster velocities \citep{Zuccarello2009}. These structures often appear in opposite bipolar pairs, though one pair may be less visible or undetected \citep{Ryutova2018}. \cite{Chen2015} highlighted that converging MMFs might be associated with flux cancellations, and thereby, triggering recurrent jet eruptions near ARs. 

In this letter, we investigate the photospheric drivers of recurrent jets observed in the periphery of active region NOAA AR 12715. We examined key magnetic field and derived parameters, that include, dynamics of fragmented flux structures or MMFs, vertical component of Lorentz force ($F_z$), and the horizontal magnetic field component ($B_h$). These parameters collectively provide critical insights into the magnetic flux transport, force balance, and shear-driven reconnection processes. The coronal extension of the magnetic field at the site of recurrent jets has been investigated using Non-Linear Force-Free Field (NLFFF) extrapolations of photospheric vector magnetograms. These observational and modeling efforts have been further assessed using 3D Radiative Magnetohydrodynamics (RMHD) Bifrost simulation, which helps to diagnose the various stages of photospheric flux cancellation leading to jet formation in the lower solar atmosphere.
\begin{figure*}[htbp]
    \centering
    \includegraphics[width=\linewidth]{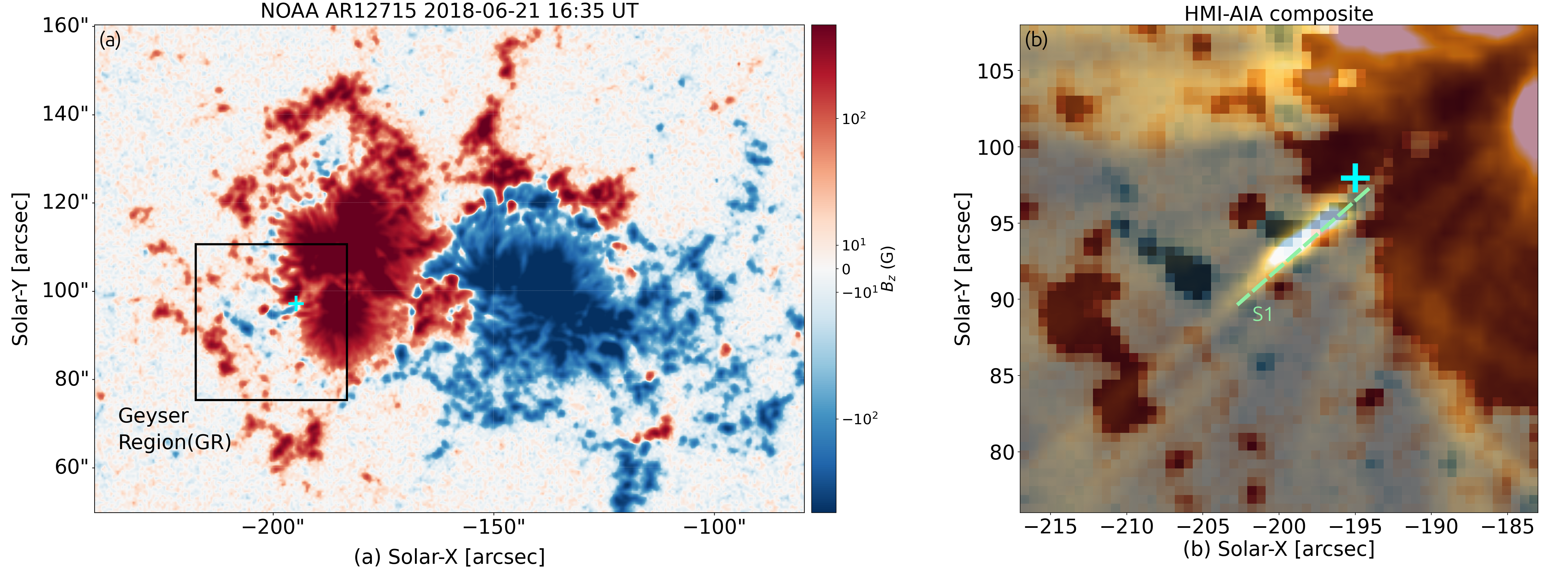}
    \vspace{0.5cm} 
    \includegraphics[width=\linewidth]{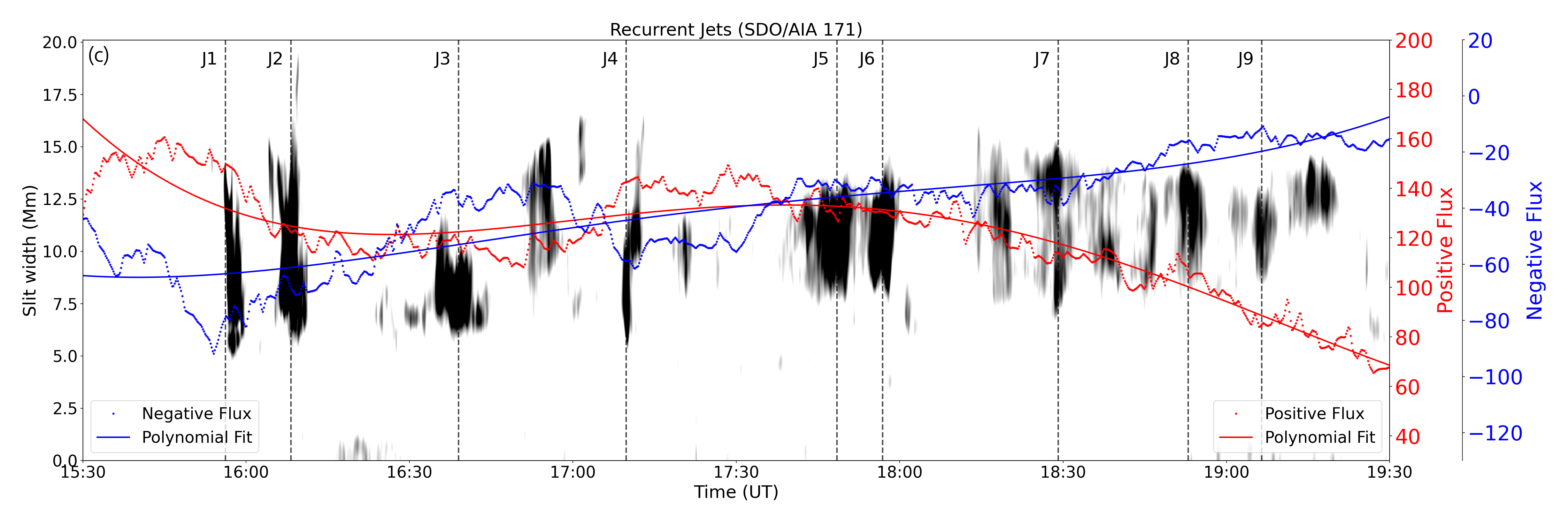}
    \caption{Panels show an overview of NOAA AR 12715 highlighting the recurrent jet–producing Geyser Region (GR). (a) Photospheric magnetic field distribution from the SDO/HMI LOS magnetogram, with region enclosed in black square highlighting the Geyser Region (GR) located at the south-eastern penumbral boundary of the active region. (b) Zoomed-in view of GR overlaid with AIA 171\,\AA\ emission, showing fragmented magnetic patches along with a jet observed at 16:35 UT. Cyan “+” symbols mark the likely footpoint of the recurrent jets, and the location of the artificial slit (S1) is indicated in lime. (c) Temporal evolution of the photospheric magnetic field in GR from 15:30 to 19:30 UT, with averaged magnitudes plotted alongside 4th-order polynomial fits. The time–distance evolution of jet activity along slit S1 is overlaid on the magnetic field evolution. Vertical dashed lines denote the onset times of the individual jet events (J1–J9).}
    \label{fig1}
\end{figure*}

\section{Observational Data and Methods}\label{data}
\subsection{Observational Data}\label{obsdata}
The data used for the analysis here are primarily from the Atmospheric Imaging Assembly \citep[AIA;][]{lemen2012} and the Helioseismic and Magnetic Imager \citep[HMI;][]{Scherrer2012} onboard the \textit{Solar Dynamics Observatory} \citep[SDO;][]{pesnell2012}. The AIA instrument provides full-disk solar images up to 1.3 R$_{\odot}$ with a temporal cadence of 12\,s and a spatial resolution of 0.6$\arcsec$. A series of recurrent jets emanating from the periphery of NOAA AR 12715 was identified using the chromospheric and coronal AIA passbands at 304\,\AA\ 
(He {\sc ii}, T~$\sim$~0.05\,MK) and 171\,\AA\ 
(Fe {\sc ix},
T~$\sim$~0.7\,MK).

HMI vector magnetograms obtained in the Fe\,\textsc{I} 6173\,\AA\ line were used to study the underlying photospheric magnetic field evolution associated with these jets. The vector field data were taken from the Spaceweather HMI Active Region Patch \citep[SHARP;][]{Bobra2014} series, which provide magnetic field estimates derived assuming a Milne-Eddington atmosphere and remapped to a Lambert cylindrical equal-area (CEA) projection.

\subsection{Photospheric Magnetic Parameters and NLFFF Extrapolation}\label{nlfff}
The HMI vector magnetograms were used to derive key parameters, including the horizontal magnetic field (B$_{h}$) and the associated vertical Lorentz force (F$_{z}$). To investigate the three-dimensional magnetic configuration at the jet source region, nonlinear force-free field (NLFFF) extrapolations were performed using the NLFFF code based on the MHD relaxation method developed by \cite{Inoue2014,Inoue2016}. The method solves zero $\beta$ MHD equations and iteratively adjusts the magnetic field toward a divergence free and force-free state consistent with the observed photospheric magnetic field. To minimize numerical errors arising from ${\bf \nabla}\cdot {\bf B}$, the hyperbolic divergence cleaning scheme of (\citealt{Dedner2002}) is employed. 

\subsection{Bifrost simulation}
Observations and magnetic field extrapolations at the sites of recurrent jet activity were compared with results from the 3D Bifrost simulation \citep{Gudiksen2011}, which is an advanced RMHD code developed to study the solar and stellar atmospheres. The computational box extends from 2.5 Mm below the photosphere to 14 Mm above it, spanning a horizontal area of 24 × 24 Mm. This domain is resolved with a high-resolution grid of 504 × 504 × 496 cells. The horizontal resolution is 48 km, while the vertical resolution varies from 20 km in the lower atmosphere to approximately 100 km in the corona.

Bifrost includes radiative transfer processes relevant to optically thick plasma in the lower solar atmosphere (photosphere and chromosphere), along with radiative losses from the lower chromosphere to corona, across the transition region. Thermal conduction is implemented along the magnetic field lines in the simulation domain.

The simulation sets the entropy inflow at the lower boundary to maintain an effective temperature ($T_{\rm eff}$) close to the solar value of 5780 K. A horizontal magnetic field of $B_y = 3363$ G is introduced at the base of the model. After $\sim$90 minutes of solar time (i.e., physical time elapsed within the simulated solar atmosphere), this field emerges through the photosphere, causing substantial changes to the pre-existing magnetic topology. A weak, oblique background magnetic field ($\leq$0.1 G) allows emerging magnetic structures to expand quickly into the overlying solar atmosphere, while driving dynamic processes and perturbing the magnetic configuration. The atmosphere responds to this flux emergence by producing dynamic features such as Ellerman bombs, UV bursts \citep[as shown using same model by][]{Hansteen2017}, and several jet-like structures (as discussed in this study), common to the lower solar atmosphere.

\begin{figure*}
\begin{center}
\includegraphics[scale=0.28]{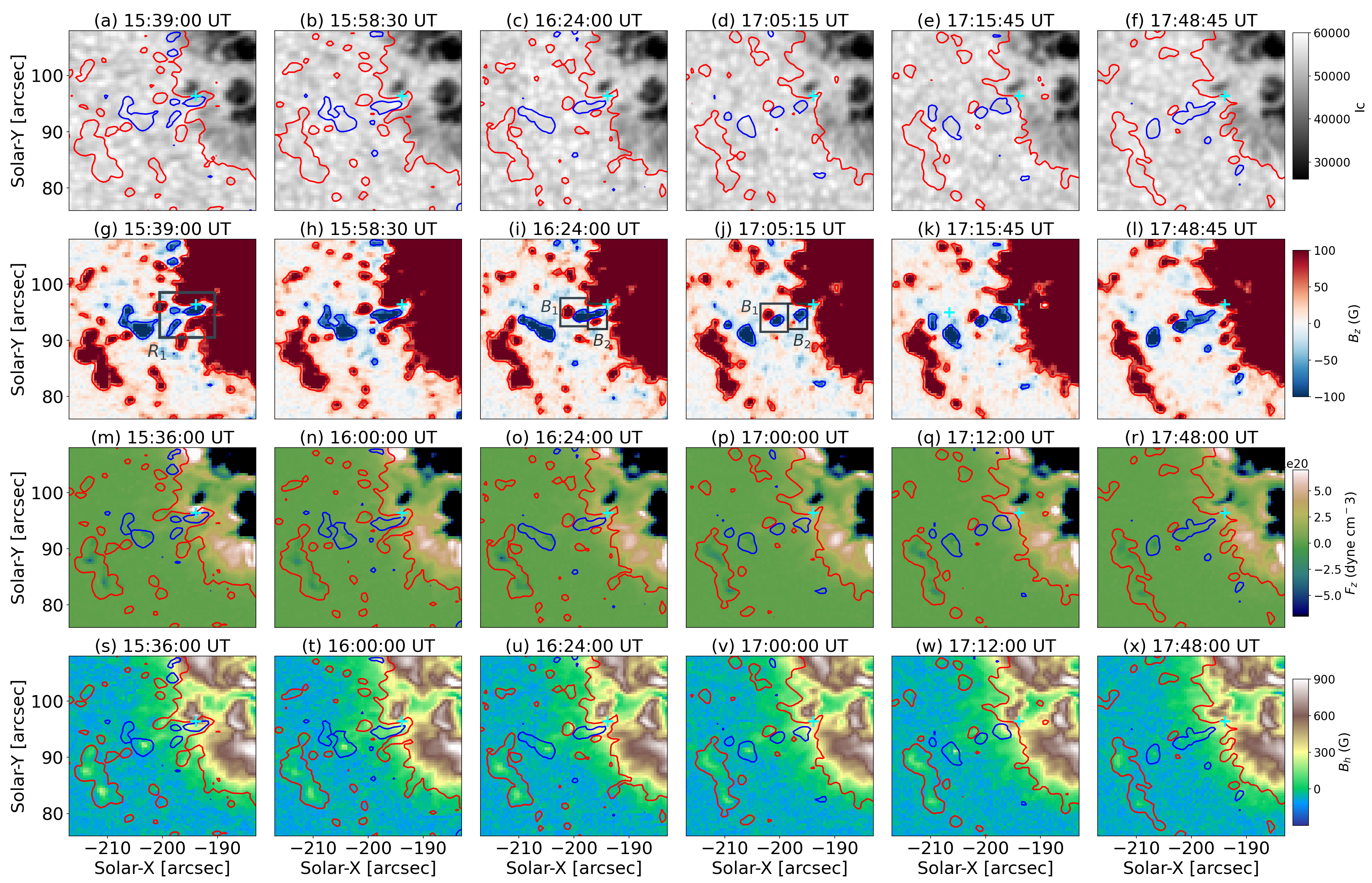}
\caption{Temporal evolution of photospheric magnetic field and its derived parameters within the GR during 15:36–17:48 UT, encompassing the period of recurrent jet activity. Panels (a–f) show HMI continuum intensity maps overlaid with contours of positive (red) and negative (blue) magnetic polarities thresholds at $\pm50$ G. Panels (g–l) show the corresponding LOS magnetic field ($B_{z}$) from HMI magnetograms. Small boxed regions, labeled `B$_1$' and `B$_2$', indicate bipolar magnetic structures with interacting polarities linked with jet eruptions. Panels (m-r) show the vertical Lorentz force component ($F{z}$), and panels (s-x) display the horizontal magnetic field ($B_{h}$). Both $F_{z}$ and $B_{h}$ exhibit noticeable variations near the recurrent jet footpoint (marked by a cyan `+' symbol). \href{run:figures/Fig2_hmi_contimuum.avi}{An animation of this figure is available}.}
\label{fig2}
\end{center}
\end{figure*}

\section{Analysis and Results}\label{results}
\subsection{Host Active region}
A series of recurrent jets were observed in SDO/AIA imaging data on 21 June 2018, originating from NOAA AR 12715, located near the disk center (N08E06). At the time, the active region was in its decay phase and exhibited a bipolar ($\beta$) magnetic configuration, with jets emerging primarily from its southeastern edge.

Figure \ref{fig1}(a) highlights the jet(s) source region (GR), characterized by fragmented magnetic features that formed small-scale bipolar configurations, creating favorable conditions for photospheric flux cancellation. This region exhibited continuous magnetic flux emergence and cancellation, with magnetic elements displaying notable outward motion. These moving magnetic features (MMFs) drifted southward, forming transient bipoles that contributed to the restructuring of the magnetic field topology in the lower solar atmosphere. The host sunspot had a wedge-shaped moat region where opposite-polarity fields emerged and appeared to trigger recurrent jet activity. Figure \ref{fig1}(b) shows a zoomed-in view of the GR with a coronal jet overlaid on the line-of-sight (LOS) magnetic field map. The jet footpoint at the wedge-shaped boundary is marked with a `+' symbol. Over the four-hour interval from 15:30 to 19:30 UT, a total of nine jet events (J1-J9) were observed originating from this region.

The time-distance plot in Figure \ref{fig1}(c) shows the episodic eruptions of recurrent jets, overlaid with the evolution of the photospheric positive and negative magnetic flux at GR. The magnetic flux are fitted with a 4$^{\text{th}}$-order polynomial to highlight the overall trend in the evolution of the photospheric magnetic fields during the period of recurrent jet activity. Both positive and negative fluxes exhibit a gradual decrease in magnitude over time. The vertical marked lines indicate the onset of individual jet events, along with the corresponding magnetic flux evolution during each episode. 

\subsection{Photospheric drivers of recurrent jet activity}
To investigate the mechanisms driving recurrent jet activity in the GR, we analyzed a combination of magnetic (e.g., flux emergence and cancellation), kinematic (MMFs), and derived photospheric parameters, including the horizontal magnetic field ($B_{h}$) and the vertical component of the Lorentz force ($F_{z}$). Variations in $B_{h}$ served as a proxy for changes in magnetic field inclination, while $F_{z}$ offered insight into shear forces that may contribute to sustained, repeated jet eruptions.

Figure \ref{fig2} shows the temporal evolution of the photospheric magnetic field and associated parameters in the GR during 15:36--17:48 UT. A sequence of HMI continuum images with overlaid magnetic flux contours (Figure \ref{fig2}(a-f)) reveals the emergence of negative polarity flux (at $\sim$ 15:39 UT) near the penumbral boundary of the active region. The wedge-shaped moat region, identified with the footpoint of recurrent jets (marked with a `+'), borders a small pore structure that gradually decreases in intensity over time. Simultaneously, LOS magnetograms (B$_z$; Figure \ref{fig2}(g-l)) show the outward drift of the emerging negative flux, leading to the formation of two bipolar structures (B$_1$ and B$_2$) through interactions with adjacent positive-polarity elements during 16:25--17:05 UT. These newly formed bipoles act as the footpoints for subsequent jets (J2-J9) from the GR. The localized evolution of magnetic flux distribution near the active region coincide with significant variation in magnitudes of both $F_{z}$ (Figure \ref{fig2}(m-r)) and $B_{h}$ (Figure \ref{fig2}(s-x)). In particular, the sequence of recurrent jet eruptions can be broadly categorized into two phases: the first jet (J1) originated from a relatively stationary site as a result of combined magnetic flux emergence and cancellation, whereas the subsequent jets (J2–J9) were associated with moving/interacting magnetic flux patches (corresponding to MMFs) that developed and evolved following the onset of J1.

\subsubsection{Jet associated stationary magnetic flux}
Recurrent jet activity near the penumbral boundary of the active region was concentrated in the GR, where fragmented magnetic field structures created a dynamic photospheric environment. The first jet in the sequence (J1) was preceded by the emergence of negative polarity flux within a wedge-shaped moat region at the penumbral boundary. Between 15:30-15:45 UT, this negative flux increased in strength from $-46\pm4$ to $-65\pm6$ G, while the surrounding positive polarity flux remained relatively constant. The newly emerged flux contributed to a buildup of magnetic energy at the jet footpoint (Fig.~\ref{fig2}(g), marked by ‘+’). Around 15:45 UT, the emerging flux began to interact with the ambient penumbral fields of opposite polarity, initiating a cancellation process that reduced the local positive flux from $183\pm17$ to $168\pm15$ G. These coupled signatures of flux emergence followed by cancellation are consistent with previous observations of reconnection-driven jet formation \citep{Chen2015, Paraschiv2020, Zhou2025}.

During this interval (15:36--17:48 UT), the vertical Lorentz force ($F_z$; Figure \ref{fig2}(m-r)) showed a notable drop in the magnitude of positive component, from $\sim9.6\times10^{20}$ to $4.5\times10^{20}$ dyne cm$^{-3}$ by 16:00 UT. This decrease reflects a rapid release of magnetic stress, likely associated with reconnection-driven plasma acceleration in the upward (or anti-clockwise) direction \citep{Kitiashvili2012, Liu2016}. Further, the horizontal magnetic field ($B_h$) near the jet footpoint decreased modestly from $822\pm80$ to $770\pm68$ G following the reconnection event (B$_h$; Figure \ref{fig2}(s-t)), suggesting a reduction in the azimuthal magnetic field component as magnetic connectivity was restructured.

\begin{figure*}[htbp]
    \centering
    \includegraphics[width=\linewidth]{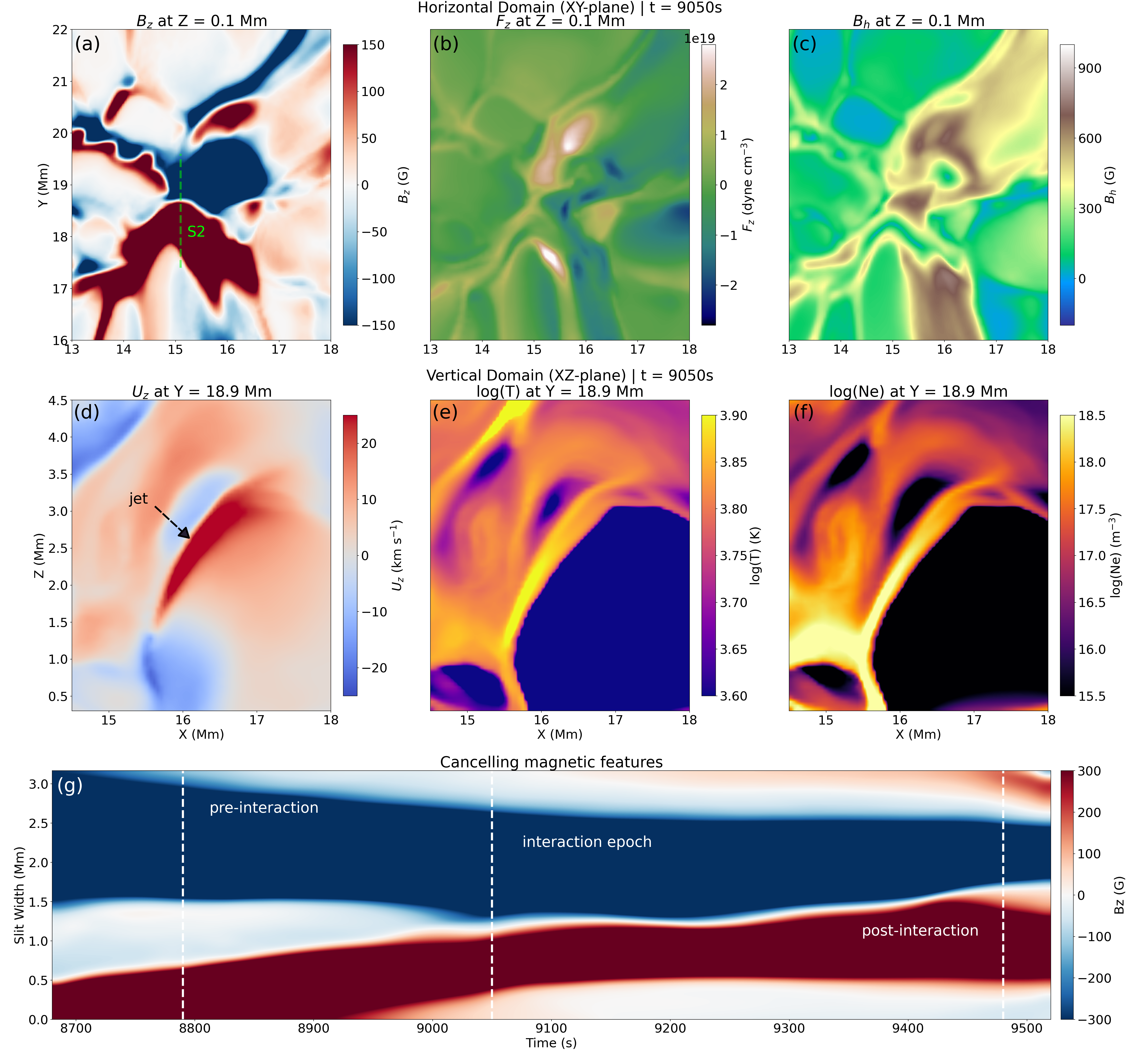}
\caption{Panels show Bifrost simulation results highlighting a photospheric bipolar magnetic feature and its associated jet. (a)-(c) Line-of-sight magnetic field ($B_{z}$), horizontal magnetic field ($B_{h}$), and vertical Lorentz force ($F_{z}$) at the photosphere ($Z = 0.1$ Mm) in the simulation domain. (d)-(f) Plasma properties of the erupted jet at $Y = 18.9$ Mm: LOS outflow velocity ($U_{z}$), temperature ($T$), and electron density ($N_{e}$) at logarithmic scale. (g) Time-distance plot along slit S2 (marked in panel a) showing the evolution of the bipolar magnetic features. The three-step process of jet triggering is evident: flux convergence ($t = 8790$ s), interaction ($t = 9050$ s), and post-interaction phase ($t = 9480$ s). \href{run:figures/Fig3_bifrost_R1.mpeg}{An animation of this figure is available}.}
\label{fig3}
\end{figure*}

\subsubsection{Moving magnetic feature(s) associated recurrent jets}
Following the onset of jet J1 at $\sim$15:56 UT, the negative polarity flux began to drifting outward from the wedge-shaped moat region, moving away from the penumbral boundary of the active region. During this process, the negative flux patch elongated from its initial compact shape, resulting in its leading-edge encountering a nearby positive polarity element. This interaction led to the formation of a new bipole, B$_1$, around 16:24 UT at (X,Y) = (-200\arcsec, 95\arcsec), see Figure~\ref{fig2}(i). The resulting MMF structure (B$_1$) migrated southward and maintained coherence until $\sim$17:05 UT (Figure~\ref{fig2}(j)). Meanwhile, the trailing edge of the same negative flux patch interacted with another positive polarity concentration, forming a second MMF bipole, B$_2$, near (X,Y) = (-196\arcsec, 94\arcsec). 

A distinct aspect of recurrent jet generation in GR is the role of MMF bipoles, which facilitated the three-step jet formation process involving flux convergence, cancellation, and subsequent outflow. The average speed of these MMFs in GR is $\sim$ 0.3-1.0 kms$^{-1}$ consistent with the previous studies \citep{Chen2015, Paraschiv2020} and slightly exceeding the average background flow speed ($\sim0.4$ kms$^{-1}$) estimated using Fourier local correlation tracking \citep[FLCT;][]{Fisher2008}. The two MMF bipoles (B$_1$, B$_2$) served as the footpoints for jet eruptions (except J1), with B$_1$ driving J3 and B$_2$ associated with J2 and J4-J5. Rest of the jet features were associated with remaining magnetic flux from B$_1$ and B$_2$ MMFs.

The negative polarity of the trailing bipolar MMF (B$_2$) interacted with a nearby positive polarity patch during its growth phase, triggering the second jet (J2) at $\sim$16:06 UT. A similar interaction involving the leading bipolar MMF (B$_1$) led to the formation of the third jet (J3) around 16:35 UT. By this time, both bipolar MMFs were fully developed. The positive polarity elements within these MMFs continued to evolve, and the subsequent fragmentation of the positive flux patch associated with B$_2$ by 17:09 UT coincided with the onset of Jet J4. The negative polarity of B$_2$ also fragmented by $\sim$17:15 UT and eventually merged with the leading MMF B$_1$. Even after the disintegration of B$_2$, ongoing interactions among the remaining opposite-polarity flux fragments sustained magnetic activity, producing additional jets (\textit{e.g.}, J5 at 17:42 UT, J6 at 17:54 UT, J7 at 18:29 UT, J8 at 18:51 UT, and J9 at 19:05 UT).

During the localized flux cancellation events linked to the MMFs, the average negative flux in the GR decreased from $-64\pm7$ to $-39\pm3$~G, while the positive flux decreased from $176\pm14$ to $153\pm15$~G. Concurrently, the magnitude of $F_z$ at the jet footpoints was higher relative to the surrounding region ($\sim 4.47\times10^{20}$~dyne cm$^{-3}$), indicating localized plasma upflows driven by magnetic pressure gradients. At the same location, $B_h$ remained high ($\sim 700\pm68$ G to $533\pm56$ G), signifying sustained magnetic shear at the footpoints. Over time, $B_h$ gradually weakened, consistent with the progressive release of magnetic stress through successive jet eruptions (Fig. \ref{fig2}(t)-(x)).

\begin{figure*}
\begin{center}
\includegraphics[width=\linewidth]{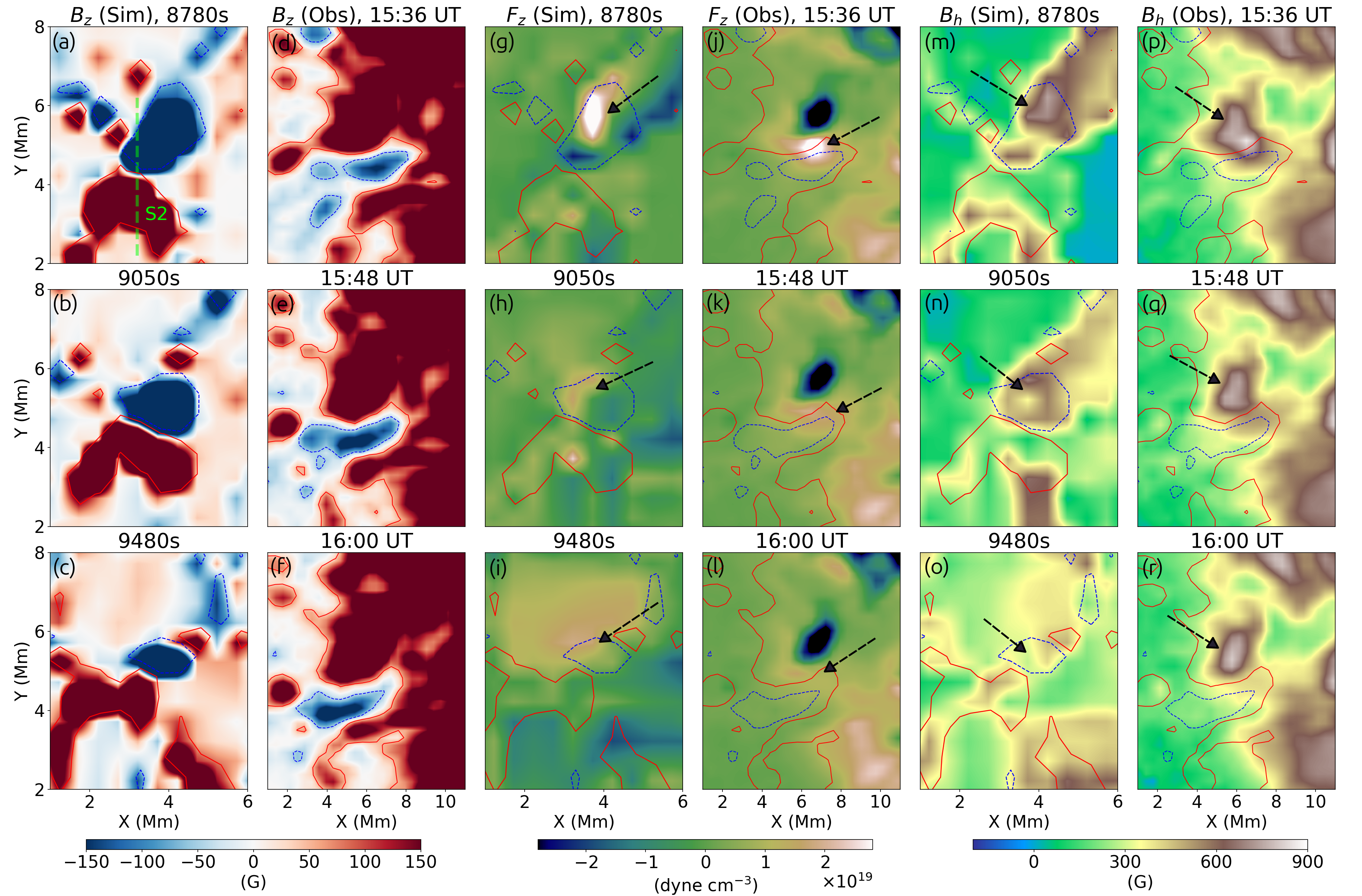}
\caption{Comparison between numerical simulation (Bifrost, resampled to HMI resolution) and observational data of the photospheric magnetic field and derived parameters, demonstrating the three-step process leading to jet formation. Panels (a)-(f) show LOS magnetic field ($B_{z}$) maps highlight the evolution of interacting bipolar magnetic patches at the jet footpoint. Slit S2, marked in panel (a), indicates the position used on the downsampled LOS magnetic field ($B_{z}$) map for correlation with Figure~\ref{fig3}(g). Panels (g)-(l) depict the vertical component of the Lorentz force ($F_{z}$), revealing the disappearance of its positive polarity at the jet footpoint prior to eruption. Panels (m)-(r) display the enhanced concentrations of horizontal component of the photospheric magnetic field ($B_{h}$), with a gradual decrease in magnitude near the flux cancellation site in both simulation and observations.}
\label{fig4}
\end{center}
\end{figure*}

\subsection{Comparison with Bifrost simulation}
The three-step mechanism underlying MMF-driven recurrent jet generation is further investigated using 3D radiative MHD Bifrost simulation that replicate the plasma and magnetic field conditions of the lower solar atmosphere. We analysed the evolution of key magnetic parameters, such as, LOS magnetic field ($B_z$), vertical Lorentz force ($F_z$), and horizontal magnetic field ($B_h$), at a height of 0.1 Mm in the simulation domain to assess their role in MMF-driven magnetic reconnection, as observed in SDO/HMI data. A case of converging and interacting bipolar magnetic patches, analogous to observational features, emerges at simulation time \( t = 8500\,\text{s} \), with significant interaction between opposite-polarity flux structures beginning around \( t = 9000\,\text{s} \). This marks the early phase of reconnection, similar to that observed in the photosphere. Figure~\ref{fig3}(a)-(c) show snapshots of \( B_z \), \( F_z \), and \( B_h \)  in the XY-plane at \( t = 9050\,\text{s} \), highlighting the onset of flux interaction.

The time-distance evolution (Fig.\ref{fig3}(g)) of bipolar magnetic patches is analyzed using an artificial slit (S2) on $B_{z}$ estimates (Fig. \ref{fig3}(a)), where these oppositely polarized flux structures tend to converge towards each other at $\sim$8780\,\text{s} (in simulation time), resulting in direct interaction at \( 9050\,\text{s}\). The interaction phase lasted approximately 480\,\text{s}, consistent with observational timescales of 420-540\,\text{s} for MMF-driven jets. Figure \ref{fig3}(b)-(c) show spatial distribution of $F_z$ and $B_h$ parameters at the photosphere with enhanced magnitudes concentrated at the site of magnetic flux interaction akin to the observations. The resulting jet signatures are evident in the simulated vertical velocity ($U_z$), temperature (log($T$)), and electron number density (log($N_e$)) (Fig.~\ref{fig3}(d)-(f)), sampled in the XZ-plane at Y = 18.9 Mm. The simulation indicate that the jet originated at a height of approximately 1.2 Mm above the photosphere, with upward plasma velocities reaching \(\sim25\,\text{km\,s}^{-1} \), and associated thermodynamic parameters of \( \log(T) \approx 3.9 \) and \( \log(N_e) \approx 18.5\).

Key photospheric magnetic field ($B_z$) and derived parameters ($F_z$, $B_h$) from simulation were also compared with observations at three distinct stages of jet formation: flux convergence, flux cancellation, and the post-reconnection phase. Figure \ref{fig4} show Bifrost simulation outputs resampled to match the HMI spatial resolution of \(0.5^{\prime\prime}\) at \( t = 8780\,\text{s} \), \( 9050\,\text{s} \), and \( 9480\,\text{s} \) with corresponding SDO/HMI observations at 15:36 UT, 15:48 UT, and 16:00 UT, respectively. These times represent the pre-interaction (flux convergence), interaction (flux cancellation), and post-interaction stages of opposite-polarity flux evolution in the photosphere.

Figures \ref{fig4} show evolution of opposite polarity LOS $B_{z}$ flux structures in simulation (panels (a)-(c)) and observations (panels (d)-(f)), saturated at $\pm$150 G. In both datasets, the flux patches decrease in size and weaken in magnitude from the pre- to post-interaction phases, consistent with photospheric magnetic flux cancellation scenario. The flux elements also elongate and fragment, producing smaller magnetic structures that can serve as recurrent jet footpoint(s). The vertical Lorentz force ($F_{z}$ ) component in both simulation and observation exhibit localized concentrations of co-spatial strong positive magnitudes with the negative-polarity magnetic patch prior to reconnection. During the flux cancellation stage and afterward, this positive $F_{z}$  signature vanishes (arrows in Fig. \ref{fig4}, panels (g)-(i) and (j)-(l)) in both datasets, though a weaker negative $F_{z}$ concentration persists in the observations.

Similarly, the horizontal magnetic field ($B_h$) shows enhanced magnitude at the flux interaction site in both simulation and observations. In the observations, $B_h$ decreases from 822$\pm$80 to 770$\pm$68 G during flux cancellation, suggesting partial relaxation of the field. In the simulation, this high magnitude feature disappears entirely after cancellation. This discrepancy between simulation and observation regarding persistent $F_{z}$ and $B_{h}$ indicate a possible role of these parameters in the generation of recurrent jets. It must, however, be noted that the studied case of interacting photospheric bipoles in simulation generated a single jet in contrast to the penumbral site in observations that produced multiple recurrent jets. 

Moreover, there were also noticeable differences between the magnitudes of the observed and simulated photospheric magnetic fields, as well as the derived parameters associated with small-scale MMFs. The average $B_{z}$ for the interacting bipolar flux patches in the observations was around 45 G, whereas the corresponding average value in the simulation was approximately 200 G. These discrepancies were reflected in the derived parameters as well, with the observed values being substantially lower than those obtained from the simulation. It is important to note that the stronger vertical magnetic fields in the simulation arise from the idealized conditions of the model, where the system evolves under controlled physical parameters without instrumental limitations. In contrast, the observed magnetic fields are obtained from vector magnetograms that are affected by instrumental uncertainties, noise, and inversion assumptions, all of which can reduce the apparent field strength \citep{Gosic2025}. Despite these differences, the simulation successfully reproduces the three-stage sequence of flux convergence, interaction, and post-reconnection relaxation, strengthening the proposed physical scenario for MMF-driven recurrent jets.

To quantify parameter variations in both the simulation and observational datasets, we computed the ratio and corresponding percentage change in their magnitudes between the pre- and post-interaction stages (Table \ref{tab1}). For the LOS magnetic field component ($B_{z}$), the observed ratio of post- to pre-interaction values is 0.757, indicating a $\sim$25\% decrease due to photospheric flux cancellation. The analogous process in the Bifrost simulation (degraded to HMI resolution) yields a ratio of 0.643, corresponding to a $\sim$36\% reduction in LOS flux. In contrast, the horizontal magnetic field ($B_{h}$) at the jet footpoint shows only a modest change in the observations, with a ratio of 0.936 ($\sim$6.4\% decrease), whereas the simulation exhibits a much larger change, with a ratio of 0.68 ($\sim$32\% decrease). This discrepancy likely arises from the persistence of the $B_{h}$ enhancement at the flux cancellation site in the observations, which is absent in the simulation. For the vertical Lorentz force ($F_{z}$), the changes are more consistent between the two cases, with ratios of 0.33 (67\% decrease) in the observations and 0.38 (62\% decrease) in the simulation. The similarity in $F_{z}$ reduction, despite the presence of concentrated negative $F_{z}$ magnitudes in the observations, may be explained by the broader distribution of weak $F_{z}$ magnitudes surrounding the cancellation site in the Bifrost simulation, which also contributes to the net change in the Geyser region.

\begin{table*}[ht]
\centering
\begin{tabular}{lcccccc}
\hline
\hline
 &  & Ratio (R) &  & & \% change &\\
\cline{2-7}
Parameters & Observations & Bifrost & Bifrost & Observations & Bifrost & Bifrost \\
 & (HMI) & (Original Res.) & (HMI Res) & (HMI) & (Original Res.) & (HMI Res) \\
\hline
${B_z} $ & 0.757 & 0.657 & 0.643 & 24.3 & 34.3 & 35.7 \\
${B_h} $ & 0.936 & 0.64 & 0.68 & 6.4 & 36 & 32\\
${F_z} $ & 0.33 & 0.39 & 0.38 & 67 & 61 & 62\\
\hline \hline
\end{tabular}
\caption{%
Variations in key magnetic parameters at pre- and post-interaction stages of magnetic bipoles associated with jet onset. LOS field strength ($B_{z}$), horizontal field strength ($B_{h}$), and vertical Lorentz force ($F_{z}$) parameters from both observations (SDO/HMI) and simulation (at original and HMI-matched resolutions). The ratio is defined as $R = X^{\mathrm{after}} / X^{\mathrm{before}}$, where $X$ denotes the respective parameter. The percentage change in magnitude is calculated as $(1 - R) \times 100$.}\label{tab1}
\end{table*}

\begin{figure*}[htbp]
    \centering
    \includegraphics[width=\linewidth]{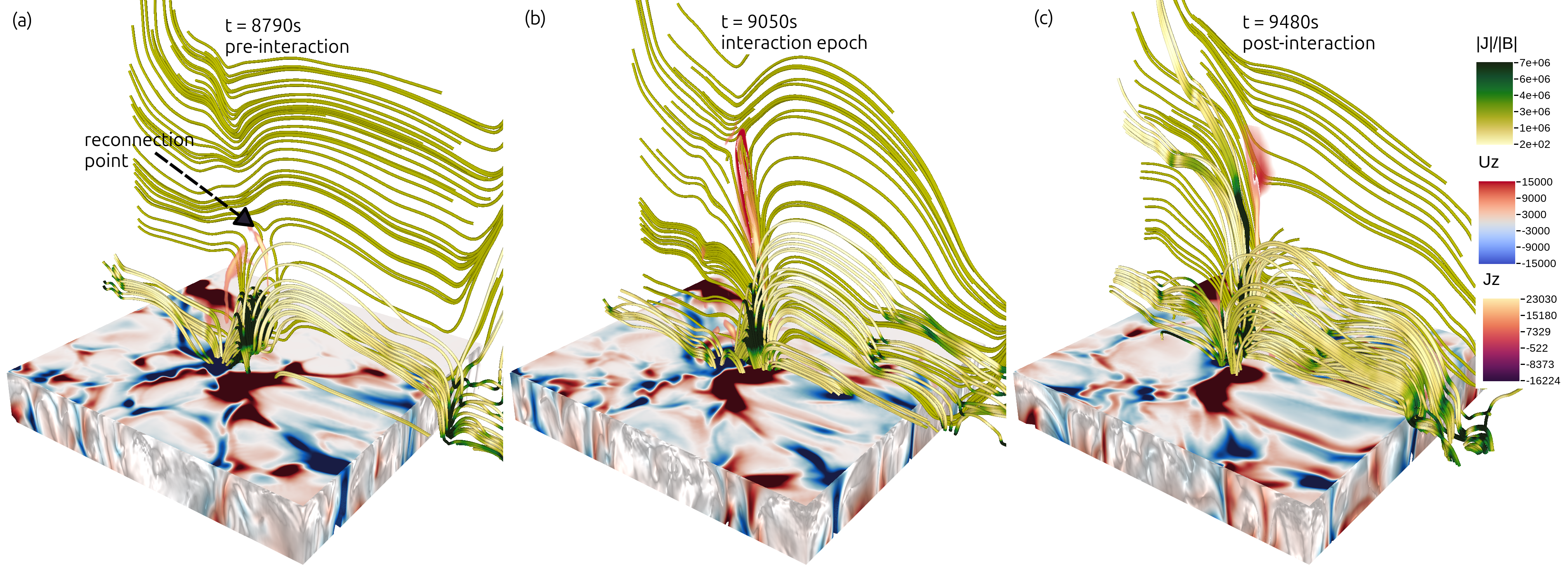}
    \vspace{0.5cm} 
    \includegraphics[width=\linewidth]{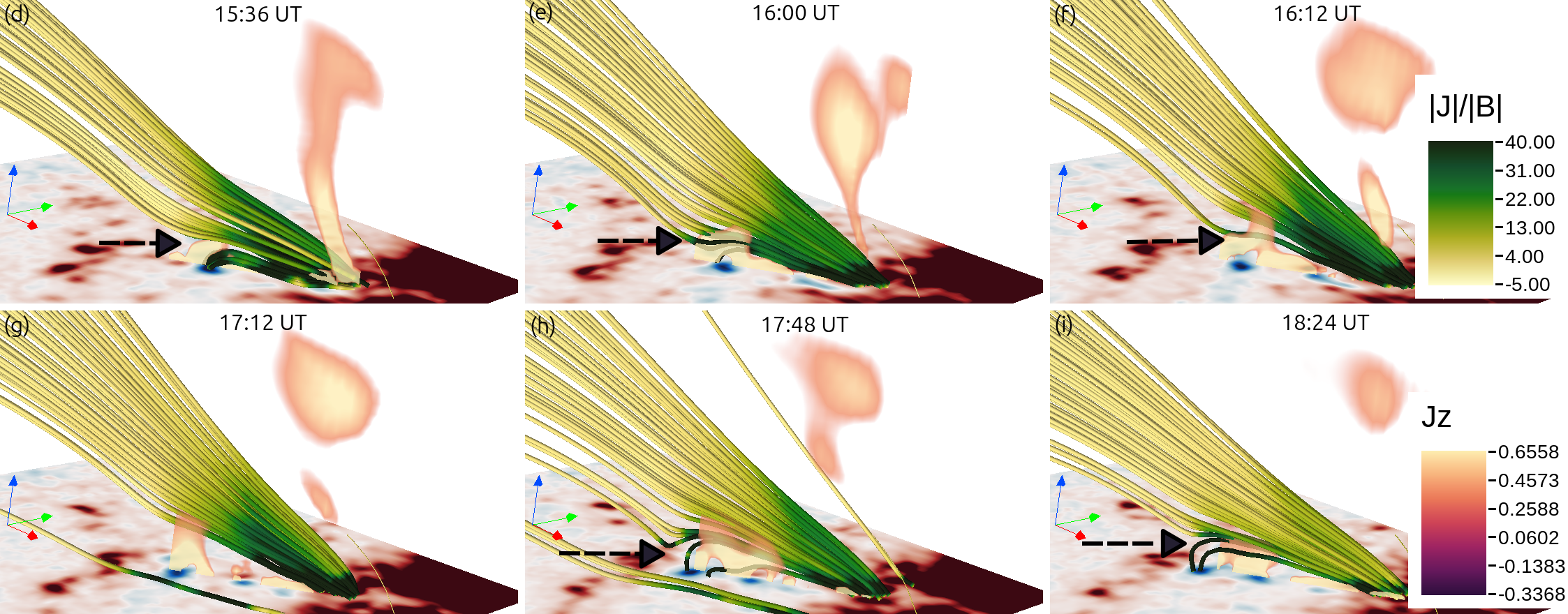}
    \caption{Magnetic field topology of the Geyser Region (GR) associated with recurrent jet activity. Panels (a)-(c) show the coronal extension of magnetic fields in the Bifrost simulation at three distinct stages of jet evolution: pre-interaction (t = 8790 s), interaction (t = 9050 s), and post-interaction (t = 9480 s). 
    Jet structures emanates from a null point formed above the interacting magnetic bipoles (indicated by an arrow in panel (a)), accompanied by enhanced current density. Panels (d)-(i) show the evolution of extrapolated magnetic field lines, revealing the formation of small-scale loop structures, associated with photospheric MMF features, beneath inclined open magnetic fields extending from the active region. The region show localized enhancements in current density (marked by an arrow) that facilitate recurrent jet formation. Both the numerical and extrapolated magnetic fields indicate the presence of magnetic dips at the sites of enhanced current density, coinciding with subsequent jet formation. \href{run:figures/Fig5_nlfff.avi}{An animation of this figure is available}.}
    \label{fig5}
\end{figure*}

\subsection{Temporal evolution of magnetic topology}
We examined the coronal magnetic field topology associated with interacting bipole patches and jet formation in both the Bifrost simulation and observations to assess the role of overlying magnetic fields in generating and sustaining recurrent jets from the Geyser Region (GR). In the observational case, the coronal magnetic field was reconstructed using nonlinear force-free field (NLFFF) extrapolations based on the magnetohydrodynamic (MHD) relaxation method \citep{Inoue2014}, which is better suited for identifying enhanced current density in a low-$\beta$ coronal environment.

In the simulation domain (Figures \ref{fig5}(a)-(c)), the jet formation follows the three-step process. During the pre-interaction phase ($t$ = 8790 s), opposite-polarity magnetic flux fragments converge, forming a magnetic null point above the interacting bipoles in the photosphere. This null point is accompanied by enhanced current density, along with bend or inward dip in overlying magnetic field lines, increasing magnetic shear at the interaction site and strengthening the current sheet, while creating conditions favorable for magnetic reconnection. In the interaction phase ($t$ = 9050 s), stressed magnetic field lines reconnect, releasing stored magnetic energy and driving plasma outflows along the magnetic dips. The rapid topological changes accelerate and heat plasma in the surrounding region. In the post-interaction phase ($t$ = 9480 s), the magnetic field gradually relaxes toward quasi-equilibrium, and the current density at the interaction site decreases noticeably. This three-stage process closely resembles the converging-field reconnection scenario proposed by \citet{Priest1994a, Dreher1997}, in which interacting flux systems produce coronal jets and localized heating.

In the observational case Figure \ref{fig5} (d-i), extrapolations reveal that GR is dominated by inclined open magnetic field structures extending from the active region into the corona. These field lines overlay mixed-polarity magnetic features in the photosphere, including MMFs that form low-lying loop structures. These loops repeatedly interact with the overlying open magnetic fields while driving continuous evolution of the local magnetic topology. Here, the overlying open magnetic field lines exhibited an inward dip or bend (Fig. \ref{fig5}(d)-(i)), resembling a kink-like instability often linked to fast reconnection across the closed–open separatrix \citep{Karpen2017}. Around $\sim$15:36 UT, the estimated vertical current density ($J_z$) increases near jet footpoints, within a region enclosed between emerging loops and the open-field system (Figure \ref{fig5}(a)). Although $J_z$ briefly declines, its magnitude rises again around $\sim$16:00 UT, coinciding with the formation and evolution of bipoles B$_1$ and B$_2$ (Figures \ref{fig5}(e)-(f)) in the photosphere. During this time, a current sheet likely forms above the PIL between bipolar MMFs leading to subsequent bending and reconnection in field lines. Variations in $J_z$ between 16:00 and 17:48 UT suggest perturbations from photospheric flux emergence and cancellations resulting in episodic reconnection between closed and open fields \citep{Archontis2010, Priest1994a}. After 17:48 UT, $J_z$ decreases substantially, indicating that the magnetic system relaxed into a more stable, lower-energy configuration with reduced reconnection efficiency \citep{Dreher1997, Archontis2010, Wyper2018}, consistent with the cessation of jet activity.

\section{Discussion and Conclusion}
In this letter, we investigate the conditions in the lower solar atmosphere that drive episodic jet activity near an active region. We focus on a case where repeated jets occurred within a mixed-polarity penumbral region of NOAA AR 12715, analyzing photospheric magnetic fields and derived parameters. To explore the mechanisms triggering and sustaining these jets in the GR, we combine multiwavelength EUV observations, 3D radiative-MHD Bifrost simulation, and magnetic field extrapolations. Our key findings are as follows:
\begin{itemize}
\item A series of nine recurrent jets (J1-J9) originated between 15:30 and 19:30 UT on 2018 May 21, from a wedge-shaped moat region at the periphery of the decaying active region NOAA 12715, characterized by fragmented magnetic polarities. The onset of first jet (J1) was associated with flux emergence near the site of gradual disintegration of a small pore, which also produced outward-propagating moving magnetic features (MMFs) at propagation velocities of $\sim$0.3-1.0 km s$^{-1}$. These MMFs (via bipoles B$_1$ and B$_2$) played a critical role in the generation of successive jet activity (J2-J9).
\item In the photosphere, MMF-driven jet recurrence were primarily governed by a three-step process: (1) convergence of opposite-polarity magnetic flux patches, (2) their interaction leading to magnetic reconnection, and (3) a post-reconnection phase involving jet plasma outflows. The outward-drifting MMFs interacted with nearby opposite-polarity flux concentrations, disrupting pre-existing magnetic topologies and driving flux cancellation.
\item These magnetic field interactions were accompanied by noticeable variations in the photospheric horizontal magnetic field ($B_{h}$) and the vertical Lorentz force ($F_{z}$) magnitudes. The evolution of these parameters was also examined using Bifrost simulation, where a converging bipolar magnetic flux patch followed the three-step process to produce a jet. Both simulation and observations indicated a decrease in $B_{z}$, $B_{h}$, and $F_{z}$ magnitudes consistent with photospheric flux cancellation. However, in the observations, both $B_h$ and $F_z$ persisted after cancellation events, suggesting their role in storing and intermittently releasing magnetic energy to power successive jets from the same site.
\item Magnetic field extrapolations revealed a complex topology with low-lying, current-carrying closed loops embedded beneath overarching open fields. This configuration is favorable for recurrent jets through episodic reconnection between the closed and ambient open fields. Following multiple eruptions, the system approached a quasi-equilibrium state in which reconnection efficiency was reduced, naturally leading to a decline in jet activity.
\end{itemize}

Our observations are consistent with previous studies \citep{Chen2015, Liu2016, Miao2019, Paraschiv2020, Murabito2021, Peng2024, Zhou2025}, which report that sites of recurrent jet activity often coincide with the gradual disintegration of a small pore or satellite spot near the penumbral boundary and/or within a wedge-shaped moat region hosting magnetic flux emergence. These studies and others (see Introduction section) largely support a scenario in which recurrent jets originate from the combined effects of magnetic flux emergence and cancellation. In this framework, flux emergence near the jet footpoints perturbs the existing magnetic field configuration, leading to flux cancellation between the newly emerged and pre-existing opposite-polarity penumbral fields \citep{Archontis2010}. While flux emergence and cancellation are widely recognized as key drivers of jet formation, the factors underlying the \textit{recurrence} of these processes may involve additional factors, such as, magnetic topology (small-scale fragmented flux structures), magnetic twist, and/or low-lying horizontal fields such as loops and arch filaments.

In our study, both observations and simulation revealed strong concentrations of horizontal magnetic fields ($B_h$) magnitude near the jet footpoints. Although this component shows a steady decline in both domains, high $B_h$ concentrations persist throughout the reconnection phase of jet formation in the observed case. Such enhanced horizontal fields indicate increased magnetic shear and presence of non-potential, current-carrying fields conducive to energy buildup \citep{Kuckein2012, Toriumi2019}. Previous studies of recurrent jets \citep[e.g.,][]{Paraschiv2020, Joshi2024} suggested generation of recurrent jets from regions having low-lying loop or mini-filament structures capable of storing and releasing magnetic energy. Our results further show a high magnitude of the vertical Lorentz force ($F_z$) in both observations and simulation, which decreases significantly ($\sim$62-67\%) from the pre- to post-jet formation stage. Such a reduction in $F_z$ is consistent with plasma ejection along open field lines, in line with scenarios where untwisting motions arise from newly emerging flux interacting with ambient open fields via interchange reconnection \citep{Pariat2009}. This emphasizes the critical role of the Lorentz force in driving plasma acceleration \citep{Kitiashvili2012, Liu2016, Yang2023}. 

An important aspect of our study is the observational evidence of MMF-driven recurrent jets. Previous works \citep[e.g.,][]{Chen2015, Paraschiv2020} also highlighted the presence of small-scale, moving magnetic fragments at recurrent jet generation site. In our case, MMFs continuously interacted with nearby opposite-polarity flux elements while driving magnetic cancellation. Using 3D radiative-MHD Bifrost simulation, we examined an analogous scenario in which converging magnetic bipoles interact and trigger a jet from a coronal null point located at a site of enhanced current density. The evolution proceeds through three stages: (i) pre-interaction stage, where newly emerged flux is advected by convective flows, building magnetic stress near null points; (ii) interaction or flux cancellation phase when opposite-polarity flux convergence initiates magnetic reconnection and plasma ejection; and (iii) post-reconnection stage when magnetic energy dissipates and field topology restructures, producing localized heating and accelerating the jet \citep{Priest1994a, Dreher1997, Priest2018}. The most likely cause for jet recurrence is the inability of the system to attain a fully stable equilibrium after the first jet eruption. Near the reconnection site, moving magnetic features (MMFs) continuously transport opposite-polarity flux and their subsequent cancellation drives repeated reconnection events. This results in a cyclic process of magnetic energy buildup and release through jet, which is then replenished by new flux inflows by MMF structures. Such a sustained flux supply may explain the observed recurrent jet activity (the ``geyser effect”), whereas the current Bifrost simulation tends to relax after producing a single eruption since there is no supply of opposite polarity flux.

Our study provides new insight into MMF-driven recurrent jet formation in decaying active regions, supported by both observations and Bifrost simulation. Upcoming observations with the Daniel K. Inouye Solar Telescope (DKIST) will deliver high-resolution spectropolarimetric measurements of the photosphere and chromosphere, enabling detailed examination of how chromospheric magnetic fields respond to photospheric dynamics driven by small-scale moving magnetic features. These estimates will enhance our understanding of magnetic stress accumulation, reconnection processes, and their role in accelerating energetic particles that contribute to the solar wind originating from solar corona.

\section*{Acknowledgments}
We thank the anonymous referees for their constructive and insightful comments. The authors thank the Solar Dynamics Observatory (SDO), a mission of NASA's Living with a Star program, and the HMI/SHARP science team for providing open access to the data used in this study. AM thanks for valuable discussions with R. Sharma and N. Liu. AM, SI, JL, and HW acknowledge support from the National Science Foundation (NSF) under grant AST-2204384. JL and HW were also supported by NSF grants AGS-2114201, AGS-2145253 (CAREER), AGS-2229064, and AGS-2309939, and by NASA grants 80NSSC20K1282, 80NSSC23K0406 and 80NSSC24K0258.

\appendix

\section{$F_z$ and $B_h$ estimations}\label{fz-bh}
\textit{For observations:} Here for HMI/SHARP-CEA data, \(F_z\) is calculated using the expression given below, $F_z = \left( B_p^2 + B_t^2 - B_r^2 \right) \Delta A$, where $B_r$, $B_p$, and $B_t$ represent the radial, poloidal, and toroidal components of the magnetic field, respectively \citep[see,][]{Sun2014}. Also, the horizontal magnetic field strength, $B_h$, is defined as $B_h = \sqrt{B_p^2 + B_t^2}$. Here, $\Delta A$ is the area of a single HMI pixel, computed using the pixel size of $0.5\,\text{arcsec}$ and the solar conversion factor $1\,\text{arcsec} = 7.25 \times 10^7\,\text{cm}$, giving $\Delta A = \left( 0.5 \times 7.25 \times 10^7 \right)^2 \approx 1.3 \times 10^{15}\,\text{cm}^2$.

\textit{For simulation:} Since the Bifrost simulation provides full access to magnetic and plasma parameters, the horizontal field is computed as $B_h = \sqrt{B_x^2 + B_y^2}$, and the Lorentz force density in SI units is calculated using the expression given below. The estimation of the Lorentz force follows the formulation described by \cite{Fisher2012}, and a similar approach has been applied in recent studies by \cite{Barczynski2019} and \citet{Maity2024}.
\[
\mathbf{F} = \frac{1}{\mu_0} \left( \nabla \times \mathbf{B} \right) \times \mathbf{B},
\]
where \( \mathbf{B} = (B_x, B_y, B_z) \) is the magnetic field vector, and \( \mu_0 = 4\pi \times 10^{-7} \, \text{N/A}^2 \) is the magnetic permeability of free space. 
The curl of the magnetic field is given as
\begin{equation}
\nabla \times \mathbf{B} =
\left(
\frac{\partial B_z}{\partial y} - \frac{\partial B_y}{\partial z},\,
\frac{\partial B_x}{\partial z} - \frac{\partial B_z}{\partial x},\,
\frac{\partial B_y}{\partial x} - \frac{\partial B_x}{\partial y}
\right).
\end{equation}

The cross product $(\nabla \times \mathbf{B}) \times \mathbf{B}$ can then be written in component form as
\begin{align}
F_x &= \frac{1}{\mu_0} \left[ (\nabla \times \mathbf{B})_y B_z - (\nabla \times \mathbf{B})_z B_y \right], \\
F_y &= \frac{1}{\mu_0} \left[ (\nabla \times \mathbf{B})_z B_x - (\nabla \times \mathbf{B})_x B_z \right], \\
F_z &= \frac{1}{\mu_0} \left[ (\nabla \times \mathbf{B})_x B_y - (\nabla \times \mathbf{B})_y B_x \right].
\end{align}

Substituting for $(\nabla \times \mathbf{B})_x$ and $(\nabla \times \mathbf{B})_y$ yields the vertical component \( F_z \) of the Lorentz force density
\begin{equation}
F_z = \frac{1}{\mu_0} 
\left[
B_y \left( \frac{\partial B_z}{\partial y} - \frac{\partial B_y}{\partial z} \right)
- B_x \left( \frac{\partial B_x}{\partial z} - \frac{\partial B_z}{\partial x} \right)
\right].
\end{equation}

Here, \( F_z \) has units of N/m\(^3\), since the magnetic field is provided in Tesla and spatial derivatives are calculated per meter. For comparison in CGS units, we converted the estimated \( F_z \) to dyne/cm\(^3\). The volume of a single pixel in the simulation domain, given a horizontal pixel size of 48 km, is calculated as \( V = (4.8 \times 10^4)^3 = 1.1059 \times 10^{14} \, \text{m}^3 \). The total vertical Lorentz force per pixel is then obtained by multiplying the vertical Lorentz force density \( F_z \) (in N/m\(^3\)) by this volume, yielding \( F_z = F_z \cdot V \) in Newtons. To convert this total force into CGS units (dynes), we use the relation \( 1 \, \text{N} = 10^5 \, \text{dyn} \). Therefore, the total vertical Lorentz force per pixel in dynes is given by
\[
F_z \, [\text{dyn}] = F_z \times (4.8 \times 10^4)^3 \times 10^5 \approx F_z \times 1.1059 \times 10^{19}.
\]
This conversion allows the direct comparison with quantities in cgs units.


\bibliographystyle{aasjournal}

\end{document}